\documentclass[twocolumn,prb]{revtex4-1}
\usepackage[T1]{fontenc}
\usepackage[utf8]{inputenc}
\usepackage{amsmath,amssymb}
\usepackage{color}
\usepackage{graphicx}
\usepackage{bm}
\begin{document}

\title{Electronic Properties of Twisted Trilayer Graphene}
\author{E. Su\'{a}rez Morell}
\affiliation{Departamento de F\'{i}sica, Universidad T\'{e}cnica
Federico Santa Mar\'{i}a, Casilla 110-V, Valpara\'{i}so, Chile}
\author{M. Pacheco}
\affiliation{Departamento de F\'{i}sica, Universidad T\'{e}cnica
Federico Santa Mar\'{i}a, Casilla 110-V, Valpara\'{i}so, Chile}
\author{ L. Chico}
\affiliation{Departamento de Teor\'{\i}a y Simulaci\'on de Materiales, Instituto de Ciencia de Materiales de Madrid (ICMM),
Consejo Superior de Investigaciones Cient\'{\i}ficas (CSIC),
C/ Sor Juana In\'es de la Cruz 3,
28049 Madrid, Spain}
\author{ L. Brey}
\affiliation{Departamento de Teor\'{\i}a y Simulaci\'on de Materiales, Instituto de Ciencia de Materiales de Madrid (ICMM),
Consejo Superior de Investigaciones Cient\'{\i}ficas (CSIC),
C/ Sor Juana In\'es de la Cruz 3,
28049 Madrid, Spain}
\date{\today}

\begin{abstract}
We study the electronic properties of a twisted trilayer graphene, where two of the layers have Bernal stacking and the third one has a relative rotation with respect to the AB-stacked layers. Near the Dirac point, the AB-twisted trilayer graphene spectrum shows two parabolic Bernal-like bands and a twisted-like Dirac cone. For small twist angles, the parabolic bands 
present a gap that increases for decreasing rotation angle. There is also a shift in the twisted-like Dirac cone with a similar angle dependence. We correlate the gap in the trilayer with the shift of the Dirac cone in an isolated twisted bilayer, which is due to the loss of electron-hole symmetry caused by sublattice mixing in the rotated geometry.  Using a tight-binding and a continuum model, we derive an effective Hamiltonian which accounts for the relevant low-energy properties of this system.
\end{abstract}

\maketitle

\section{Introduction}

The electronic properties of few-layer graphene have a remarkable dependence on the stacking arrangement. 
Bilayer graphene with Bernal or AB stacking, which is the most energetically favorable structure, has a low-energy parabolic spectrum; however, with direct or AA stacking, the carriers remain massless with a linear dispersion. Rotational disorder in the stacking has been experimentally evidenced and theoretically studied. Twisted bilayer graphene presents an intriguing low-energy behavior as a function of the rotational angle between layers, ranging from linear dispersion  with a Fermi velocity equal to that of monolayer graphene to a velocity renormalization for small rotation angles.\cite{Lopes_2007,Morell_2011b,Xian_2011,Bistritzer2011,Luican_2011,Li_2010,Lopes_2012,Trambly_2012a,SanJose_2012} 
 Applying a perpendicular electric field to these bilayer phases yields different results: while the AA or twisted bilayer graphene remains gapless, for the AB stacking there is a gap induced by an external voltage,\cite{McCann_2006b,Fai_2009,Zhang_2009} which is extremely important for device applications. 
Symmetry and changes in the interlayer interaction due to the different stacking arrangements are responsible for this variety of behaviors.\\

Trilayer graphene also presents a remarkable dependence on the stacking order. Bernal or ABA stacking shows an energy dispersion relation that looks like the superposition of the monolayer massless linear bands and the parabolic massive AB-like bands. Differently from the Bernal bilayer, no gap opens in the Bernal ABA trilayer under the application of an electric field. However, rhombohedral trilayer graphene, with ABC stacking, does present a gap with an applied electric field.\cite{Aoki_2007,Cracium_2009,KinFai_2010,Lui_2011} 
These distinct behaviors can be related to the different symmetries of the stackings: the ABA trilayer shows mirror symmetry, whereas the ABC rhombohedral trilayer does not, having instead spatial inversion symmetry, as the Bernal bilayer. 
Trilayer graphene also may present turbostratic disorder, either complete, with all layers disoriented, or partial, with two layers stacked AB and a third layer showing a relative rotation. These twisted trilayers have been experimentally found on SiC-grown samples, graphene on graphite and in exfoliated few layer graphene.\cite{Varchon_2008,Miller_2010,Trambly_2012b} Some previous theoretical works have partially addressed the properties of these twisted trilayer structures, but a detailed analysis is still lacking.\\

In this work we study a trilayer graphene composed of two layers with Bernal stacking and a third one rotated from an initial ABA position, i.e., an AB-twisted layer (ABT). We explore the dependence of the electronic properties on the rotation angle $\theta$ employing a tight-binding and a continuum model. This allows us to derive an effective Hamiltonian which describes the low energy behavior of this system.

Our main results are the following: \\
(i) The ABT graphene trilayer shows a combination of bilayer Bernal-like parabolic bands with states located on the AB-stacked layers, and linear twisted-like bands forming a Dirac cone, with states mainly found in the twisted layer. \\
(ii) The parabolic bands develop a gap that increases for diminishing angle. Likewise, a shift of the twisted-like Dirac point is observed with a similar angular dependence.  The velocity of the twisted-like bands is renormalized and tends to zero when $\theta\rightarrow 0$.\\
(iii) We find that the gap is correlated with a shift of the Dirac point of the isolated twisted bilayer. This shift is 
 related to the loss of electron-hole symmetry due to the mixing of the two sublattices produced by the twisted layer. \\
 (iv) In order to understand the band structure and analyze our results, we have compared the 
 outcome of 
 our tight-binding model with 
 that 
 attained by using a generalization of the continuum approximation of Lopes dos Santos.\cite{Lopes_2007} We obtain that the shift of the Dirac cone of the isolated twisted bilayer is erroneously described by the continuum model.
 As a consequence, all the low-energy physical properties related to this shift, such as the gap and the position of the Dirac point of the twisted-like bands in the trilayer, are not appropriately described by the continuum model. However, the velocity renormalization is suitable described by the continuum  approximation.\\
(v) Equipped with the knowledge gained from the comparison of the tight-binding and the continuum model, we derive an effective Hamiltonian which captures the low-energy physics of these trilayers, giving a simple description of their dispersion relation. \\
 

%
The paper is organized as follows: we describe the geometry of the twisted trilayer and models employed in Sec. \ref{sec:geomod}. In Sec.  \ref{sec:results} we present our results, which we discuss in the light of the continuum model. We finish with a summary of our main results in Sec. \ref{sec:sum}.

\section{Geometry and Model}
 \label{sec:geomod}
\subsection{Geometry}

We have studied a trilayer graphene composed of two layers with Bernal stacking plus a third one with an 
arbitrary relative rotational angle with respect to the other two. This twisted layer occupies an outer position in the stacking. 
In order to build a commensurate unit cell we have followed a procedure to find coincidence lattice points in the crystal similar to that described by Campanera {\it et al.} for a twisted bilayer.\cite{Campanera_2007} First, we have three AB-stacked graphene layers (ABA) and then we rotate one of the outer layers. We choose a B site, i.e., a site where an atom in one layer is exactly at the center of the hexagon of the lower layer, as our rotation center. We select a site of the layer with coordinates $\mathbf{r}=m \mathbf{a}_{1} + n \mathbf{a}_{2} $ and rotate it an angle $\theta$ to an equivalent site $\mathbf{t}_{1}=n \mathbf{a}_{1} + m \mathbf{a}_{2}$, where $\mathbf{a}_{1}=(-1/2,\sqrt{3}/2)a_{0}$ and  $\mathbf{a}_{2}=(1/2,\sqrt{3}/2)a_{0}$ are the graphene lattice vectors; \textit{n},\textit{m} are integers, and $a_{0}=2.46\,$\AA\  is the graphene lattice constant.  The unit cell vectors of the trilayer twisted cell can be chosen as $ \mathbf{t}_{1}=n \mathbf{a}_{1} + m \mathbf{a}_{2} $ and $\mathbf{t}_{2}= -m \mathbf{a}_{1} + (n+m) \mathbf{a}_{2}$. The trilayer unit cell built in this way has $N=6 (n^{2}+ mn+m^{2})$ atoms and can be labeled by the indices $(n,m)$, which identify the twisted bilayer part of the structure.\cite{Morell_2010,Trambly_2010} The distance between layers is set to $3.35$ \AA.\\

\begin{figure}[htbp]
\includegraphics[width=\columnwidth,clip]{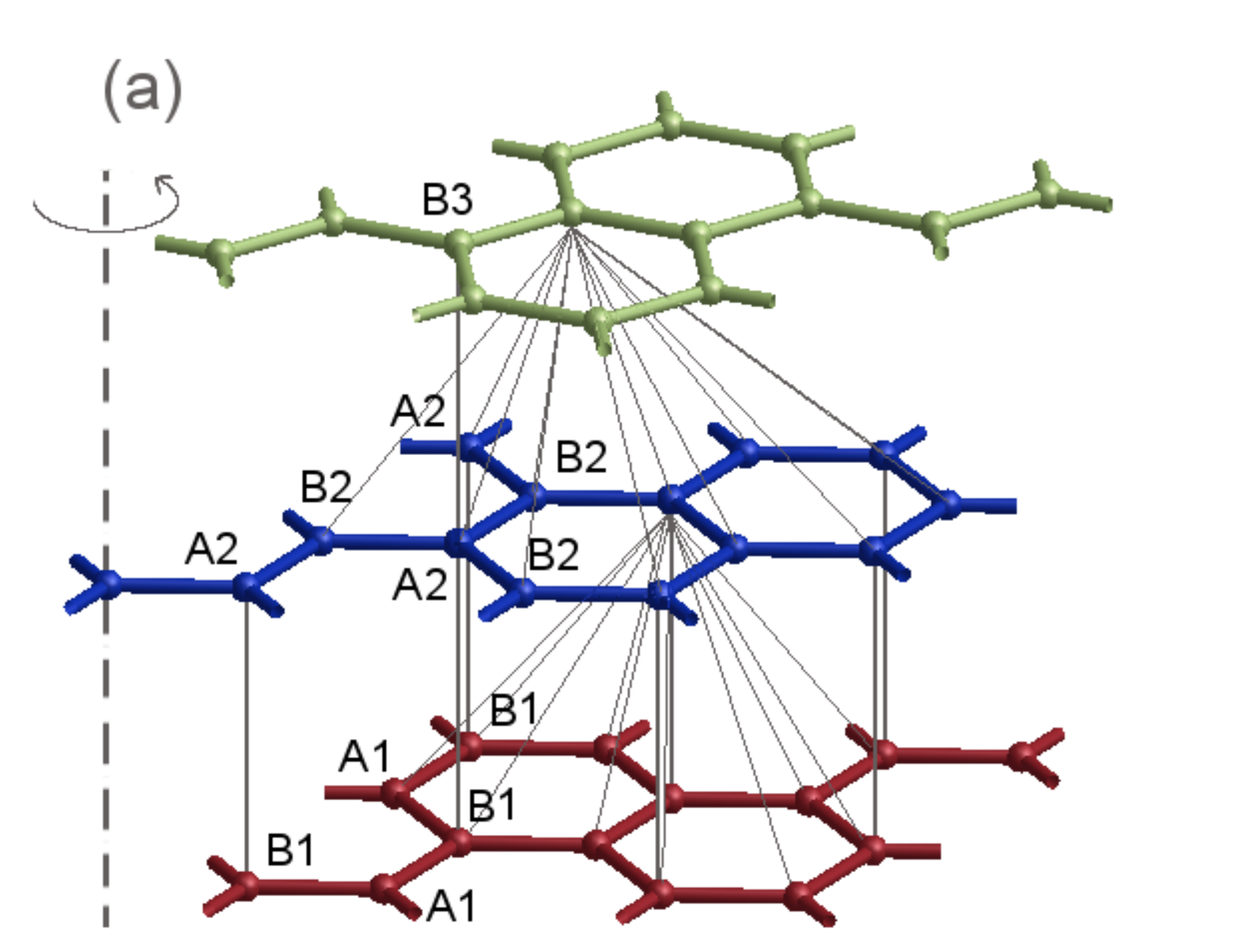}
\includegraphics[width=\columnwidth,clip]{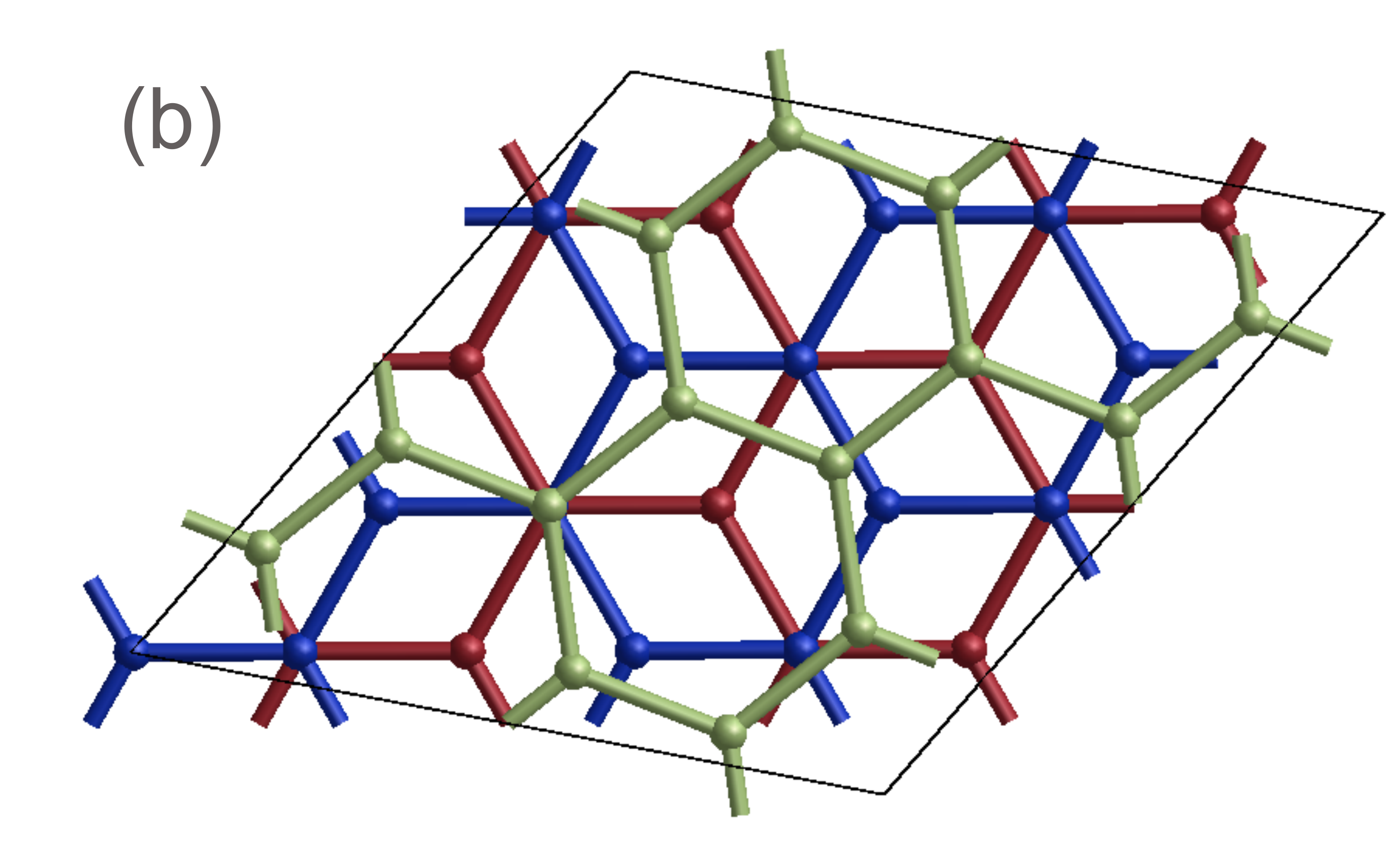}

\caption{(Color online) Unit cell for the ABT (2,1) trilayer graphene. The rotation angle $\theta$ is 28$^{\rm o}$ and the number of atoms equals 42. (a) Side view of the unit cell, indicating  the rotation axis for the twisted layer and a schematic view of the hoppings considered in the tight binding model (the chosen set of parameters implies hoppings beyond the unit cell). (b) Top view of the unit cell. The Bernal-stacked layers are shown in red and blue (dark gray) and the twisted later in green (light gray). } 
\label{fig:geom}
\end{figure}

Figure \ref{fig:geom} shows the unit cell for a (2,1) trilayer, where the top graphene sheet has been rotated with respect to the Bernal-stacked bilayer. 
This system can be 
viewed as
  two structures stacked together with a common layer, one 
  being a Bernal bilayer graphene and the other 
  a twisted bilayer graphene.\cite{Lopes_2007} 
The geometry of the twisted unit cell has been extensively studied by Shallcross \textit{et al.}\cite{Shallcross_2010} and Mele.\cite{Mele_2010}
%

\subsection{Tight Binding Model}
\label{sec:TB}
We model the trilayer graphene band structure within the tight binding approximation including only the $p_{z}$ orbitals. Within each layer, we consider a fixed nearest-neighbor intralayer hopping parameter  $\gamma_0=-3.16$ eV. We take into account the tunneling  between adjacent layers, neglecting the direct hopping between the outermost ones.\cite{notahop13} For the layer-layer interaction we employ a distance-dependent
 hopping. \cite{Shallcross_2010,Trambly_2010,Morell_2010}
  Thus, the Hamiltonian is
given by $H=H_{1}+H_{2}+H_{3}+H_{12} + H_{23}$, where $H_{n}$ ($n=1,2,3$) is  the Hamiltonian for  the individual layer  $n$ 
and $H_{mn}$ describes the interlayer coupling between consecutive layers $m$, $n$:
 \begin{equation}
H_{mn} =  \sum_{i,j} \gamma_{1}e^{-\beta (\mathbf{r_{ij}}-\mathbf{d})} c^\dagger_{i} c_{j} + H.c.,
\label{eq:ham}
\end{equation}
where $\gamma_{1}=-0.39$ eV is the nearest-neighbor interlayer hopping parameter, $\mathbf{d}$ is the interlayer distance, $\mathbf{r_{ij}}$ is the distance between atom $i$ on layer $m$ and atom $j$ on the other layer $n$, and $\beta =3$. This value of $\beta$ accurately reproduces the dispersion bands calculated within a Density Functional Theory approach.\cite{Morell_2010,Shallcross_2008,Latil_2007}
 We let every atom in a layer interact with the atoms in the adjacent layer located inside a circle of radius $6 a_{CC}$, where $a_{CC}$ is the nearest-neighbor distance between carbon atoms, equal to 1.42 \AA. This takes into account the complexity of the unit cell and at the same time breaks the electron-hole (e-h) symmetry due to the fact that we are mixing the two sublattices. 
Notice that we employ the same interlayer Hamiltonian to model the interaction between the Bernal-stacked and the twisted layer.

\subsection{Continuum Model}
\label{sec:cont}
We have developed a continuum model for describing the twisted  trilayer graphene that is a generalization of 
that 
proposed by Lopes dos Santos\cite{Lopes_2007} for twisted bilayer graphene.

The low energy properties of the $AB$ twisted trilayer graphene are described in the continuum approximation and for angles between 1$^{\rm o}$ and 20$^{\rm o}$, see Appendix, by the following $4\times4$ Hamiltonian,

\begin{equation}
H^{4  \times 4}=  \left( \begin{array}{cccc}
0 &  \frac {v_F \tilde v }{\gamma _1}  (g_+)^ 2 &0 &0 \\
\frac {v_F \tilde v }{\gamma _1}     (g^* _+)^2  &\bar {\epsilon} _0&0 &0  \\
0& 0& \bar {\epsilon} _0&  \tilde v   g_-   \\
0& 0& \tilde v    g_-^*   & \bar {\epsilon} _0\\
 \end{array} \right) 
 \label{ham_cont}
\end{equation} 
where $\tilde v$ and $\bar {\epsilon} _0$ are the renormalized velocity and the shift of the Dirac cone of the twisted bilayer respectively\cite{Lopes_2007} and 
$g _{\pm}$=$g({\bf k}\pm  \frac {\Delta {\bf K}} 2) $ with $g(\bf k)$=$k_x -i k_y $, $\Delta {\bf K}$ being the vector connecting the Dirac cones of the rotated layers.
The basis for this Hamiltonian are the non-dimer atoms $A_1$ and $B_2$ of layers $1$ and $2$ respectively, and the two basis atoms $A_3$ and $B_3$ of layer $3$.
From the diagonalization of the previous Hamiltonian we obtain the following four bands,
\begin{equation}
-\frac {v_F \tilde v }{\gamma _1}  |g_+|^ 2 \, , \,  \bar {\epsilon} _0 + \frac {v_F \tilde v }{\gamma _1}  |g_+|^ 2 \, \, {\rm and} \, \, \bar {\epsilon} _0 \pm \tilde v   | g_-|    \, .
\label{bands_cont}
\end{equation}

Notice that the continuum model describes the bands corresponding to a given valley of the original graphene layer.
 Interchanging $g _{+}$ by $g _{-}$ in  Eq. \ref{ham_cont} yields the results for the other valley, for which the position in $k$ space of the parabola and the Dirac cone are swapped. In order to compare to the TB results, the bands for both valleys should be superposed. This is equivalent to substitute $g_\pm$ by $g$ in Eq.  \ref{ham_cont}. 

\section{Results and Discussion}
 \label{sec:results}

\begin{figure}
\includegraphics*[width=\columnwidth,clip]{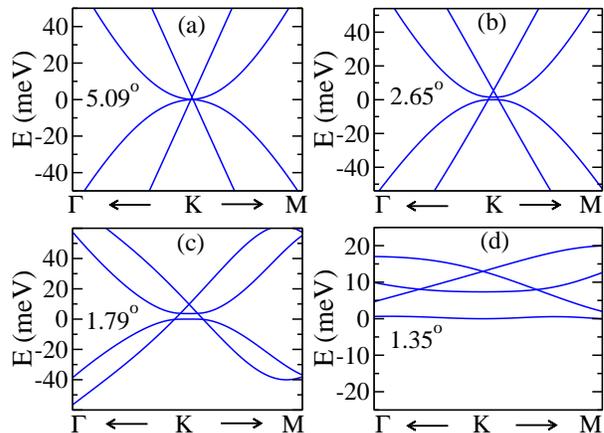}
\caption{(Color online) Band structures corresponding to four instances of ABT trilayer graphene. The corresponding angles are indicated in each panel.  } 
\label{fig:bands}
\end{figure}

Figure \ref{fig:bands} shows the tight-binding band structures for four instances of ABT twisted trilayers with angles comprised between 1.35$^{\rm o}$ and 5.09$^{\rm o}$. Two linear bands corresponding to the twisted layers and the parabolic bands stemming from the Bernal-stacked layers are clearly seen for all the cases depicted. For the largest angle shown in Fig. \ref{fig:bands} (a), 5.09 $^{\rm o}$, the four bands are practically almost degenerate at zero energy and in the scale of energies of 
the plot the spectrum looks like e-h symmetric. 
However, for smaller angles, a gap between the two parabolic bands and a shift of the Dirac point in the linear bands can be discerned. In all the 
depicted
twisted angles, the lowest energy parabolic band is pinned at zero energy. This band is very weakly coupled
and has a non-bonding character; for this reason its energy at the Dirac point does not vary with the rotation angle. The Dirac point and the minimum of the other parabolic band have 
energy shifts that do depend on the twist angle. Besides, we find that the degree of e-h asymmetry increases dramatically with diminishing angle, see  Fig. \ref{fig:bands} panels (b) to (d).

\begin{figure}[htbp]
\includegraphics[width=\columnwidth]{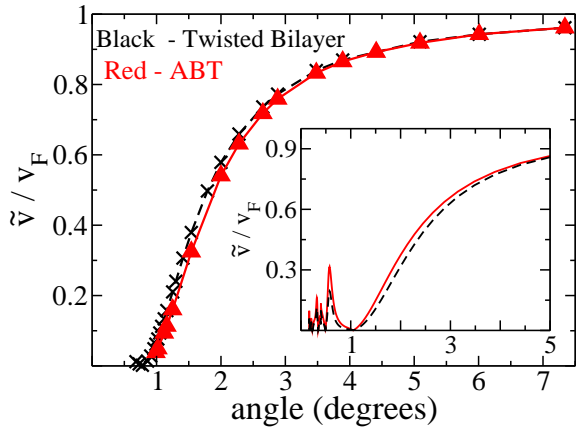}
\caption{(Color online) Fermi velocities of the trilayer graphene twisted-like bands and the twisted bilayer graphene normalized to the Fermi velocity of monolayer graphene $v_F$ calculated within the TB approximation. In the case of the trilayer, we stop the plot of the velocity at the angle 
for which 
a gap develops at the Dirac point.  Inset: Fermi velocities of the trilayer twisted-like bands and the isolated twisted bilayer calculated with the continuum model. Several zeros of the velocities can be distinguished. 
 } \label{fig:vel}
\end{figure}

{\it Velocity renormalization.} The Dirac-like bands  clearly show a Fermi velocity reduction when the twisted angle 
decreases.   
This is patent for the smaller angle depicted, 1.35$^{\rm o}$,
for which a reduction of near  90$\%$ is observed (Fig.\ref{fig:bands} (d)). 
In order to quantify the velocity renormalization, in Fig. \ref{fig:vel} we compare the velocities of the twisted Dirac-like bands in the trilayer graphene with those of the twisted bilayer case, both obtained in the tight-binding approximation. We see that the behavior of the two velocities is similar up to 1$^{\rm o}$. In both cases the band velocity is strongly renormalized for rotation angles below 5$^{\rm o}$, and tends to zero when $\theta \rightarrow 0$. 

Within our model TB Hamiltonian we find for the twisted bilayer that  the  velocity  vanishes at an angle near 0.8$^{\rm o}$, and then the velocity increases again for smaller angles. 
It has been proposed that for even smaller angles new zeros in the velocity should appear.\cite{Bistritzer2011} As in TB calculations smaller twisted angles correspond to larger unit cells, we do not find a second zero in the Fermi velocity because of the numerical limitations to the angles we can study. 
In the twisted trilayer case the situation is different: we find that for angles smaller than 1.16$^{\rm o}$, where the velocity is near zero, the conduction and valence bands couple and a small anticrossing gap of the order of  $10^{-4} \gamma_0$  appears. As we will show below, this anticrossing occurs because at small angles 
the layers are strongly coupled, resulting in wave functions with a mixed weight in the three layers. 
By restricting  the tunneling between the twisted bilayer and the extra layer only to vertical hoppings, we reduce the mixing between layers and this anticrossing disappears.

In order to analyze our results we have computed the  band structure of the trilayer twisted bilayer in a continuous model. In the inset of Fig. \ref{fig:vel}, we plot the velocity of the Dirac point obtained with the continuous model for the twisted bilayer and trilayer graphene. As in the tight-binding results, we see that the velocity renormalization in the bilayer and trilayer cases is practically the same.
The continuum approach 
allows for the exploration of smaller angles than the TB model. 
Thus, it is found that the zeros in the velocity appear at the same angle, both in the bilayer and in the trilayer cases
(see inset of Fig.\ref{fig:vel}), and they correspond to quasi-confined states in the $AA$-stacked regions of the twisted bilayer.\cite{Lopes_2012,Trambly_2012a,SanJose_2012}
Contrary to the TB results, 
there is no gap in the Dirac-like bands at small rotation angles in the continuum approximation. 
As we commented above, this is because within this approach the tunneling between layers $1$ and $2$ is treated as vertical, and this 
yields a zero gap also in the TB calculation.

\begin{figure}[htbp]
\includegraphics[clip,width=\columnwidth]{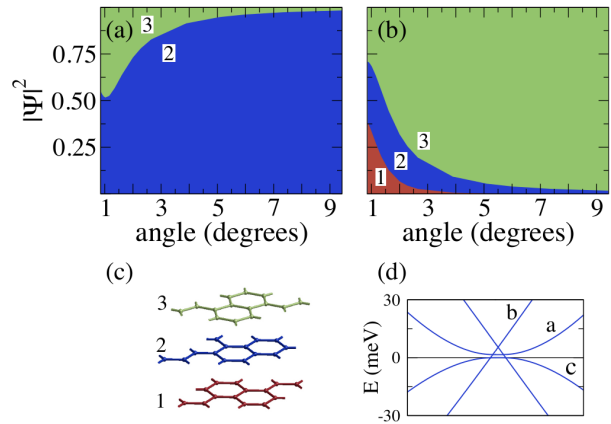}
\caption{(Color online) Electronic probability densities corresponding to the (a) parabolic and (b) linear bands distributed in the three layers. 
Red, blue and green colors and numbers 1, 2, 3 code the three layers from bottom to top, namely (1) outer Bernal layer, (2) middle Bernal layer, and (3) twisted layer, as depicted in panel (c). 
Letters {\it a, b, c} label the shifted parabolic, linear, and zero-energy pinned parabolic bands respectively, as indicated in the schematic band structure shown in panel (d). 
 } \label{fig:dens}
\end{figure}
{\it Layer distribution.} The spatial distribution of the carriers corresponding to different bands gives an idea of the degree of coupling in the system. The carriers corresponding to the 
band pinned at zero energy, labeled $c$ in Fig. \ref{fig:dens}(d), are situated in the outer Bernal layer (1), independently of the rotation angle. The behavior of the other bands is more complex. 
In panels (a) and (b) of  Fig. \ref{fig:dens} we present the electronic probability densities of the 
shifted parabolic band $a$ and of the linear twisted-like bands $b$, distributed in the three layers, see Fig. \ref{fig:dens}(d). We have plotted the probability densities of the eigenstates close to the Dirac point, checking that the two states corresponding to the Dirac cone have the same distribution. The 
shifted parabolic band $a$ corresponds to states mainly located in the inner Bernal layer, labeled 2, especially for large rotation angles.  The probability of being in the twisted layer (3) increases with diminishing angle, most notably below 5$^{\rm o}$. Likewise, the carriers corresponding to the linear twisted-like band, band $b$ in Fig. \ref{fig:dens}(d),  are mostly located in the twisted layer (3), but the probability of being in the other two layers increases below 5$^{\rm o}$, being almost equally distributed among the three layers for angles smaller than 1$^{\rm o}$. Thus, for smaller angles we obtain  that the carriers corresponding to  bands  $a$ and $b$ have an increasingly mixed spatial distribution, evidencing an enhancement of the coupling. 
This angle of 5$^{\rm o}$ is where the renormalization of the velocity for the twisted trilayer carriers is noticeable; as it is well known, this is also a sign of the increased coupling, that we also relate here to the spatial distribution of the electrons corresponding to the low-energy bands.

\begin{figure}[htbp]
\includegraphics[width=\columnwidth]{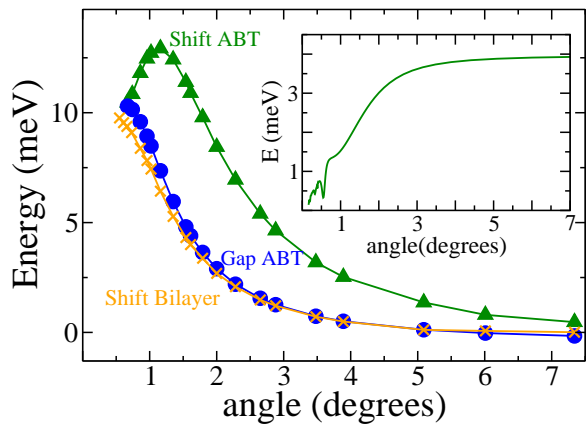}
\caption{(Color online) Shift of the Dirac cone (triangles) and gap between the parabolic bands (circles) for ABT trilayer graphene along with the shift for the Dirac cone for the isolated twisted bilayer (crosses) calculated with the TB model. Inset: Shift of the Dirac cone for the ABT trilayer obtained with the continuum model. 
 } \label{fig:sh}
\end{figure}

{\it Energy gap and shift.} In Fig. \ref{fig:sh} we show the shift of the Dirac cone and the gap between the Bernal-like parabolic bands found for the ABT trilayer calculated with the TB model. 
For comparison, we also plot the shift for the Dirac point in an isolated twisted bilayer graphene calculated with our tight-binding model. The gap increases with diminishing angle, following closely the shift of the Dirac cone for the isolated bilayer.  

In the continuum approximation, there is also 
a gap in the parabolic bands that is equal to the shift of the Dirac cone $\bar {\epsilon}_0$,  Eq. \ref{bands_cont}.
However, the 
dependence
of this gap with the rotation angle is 
at variance with 
the tight-binding results. In the inset of Fig. \ref{fig:sh} we plot 
$\bar {\epsilon}_0$ as function of the rotation angle. The shift is obtained by diagonalizing the full continuum Hamiltonian in a large enough plane wave basis, see Appendix. 
In the continuum approximation the gap 
has an almost constant value equal to  $\gamma_1 ^2 /(2 \pi \sqrt{3} \gamma_0)$ for angles in the range 4$^{\rm o}$ and 20$^{\rm o}$,\cite{Lopes_2007} and decreases to zero at smaller angles. The oscillations observed in the continuum results are related to those observed in the renormalized velocity (inset of Fig. \ref{fig:vel}). 
We attribute the discrepancy between the tight-binding and continuum results to the fact that the continuum model can only cope with the layer-layer interactions locally.
The vertical tunneling
neglects the effect of odd-numbered carbon atom rings among layers considered in a TB model with hopping beyond nearest neighbors. As a consequence, 
the continuum model 
does not appropriately describe the electron-hole asymmetry of the
system. 

Fig. \ref{fig:sh} also shows that around 5$^{\rm o}$, the two parabolic bands cross,
so for large angles the 
non-bonding band pinned at zero has a higher energy than the shifted parabolic band at the Dirac point. This change in the position of the two parabolic bands coincides with the change in the spatial distribution of the states corresponding to the linear twisted-like band and the 
shifted parabolic band. 

{\it Continuum effective model}. As the continuum model outlined in the previous Section is not valid for the description of the low-energy spectrum of twisted trilayer graphene, we modify it accordingly, deriving an effective 
Hamiltonian 
for the band structure near the Dirac point. This Hamiltonian describes the parabolic Bernal-like bands and the linear Dirac cone corresponding to the twisted layer:

\begin{equation}
H_{\rm eff} =  \left( \begin{array}{cccc}
0 & \frac{v_F\tilde{v}}{\gamma_1} k^2 & 0 &0 \\
 \frac{v_F\tilde{v}}{\gamma_1} k^2 & \epsilon_1 & 0 & 0 \\
0 & 0 &  \epsilon_0 &  k \tilde{v} \\
0 & 0  &  k \tilde{v} &  \epsilon_0
\end{array} \right),
\label{hameff}
\end{equation} \\

\noindent
is the wavevector measured from the Dirac point, 
and $\epsilon_0$, $\epsilon_1$ are small onsite energy terms that break the e-h symmetry, giving rise to the shift of the twisted cone and the opening of the gap respectively.  
This model yields 
a decoupled linear Dirac cone stemming from the twisted layer and the parabolic bands from the Bernal stacking. 
%
%

The difference between this Hamiltonian and that given by Eq.  \ref{ham_cont} amounts to the diagonal terms $\epsilon_0$ and $\epsilon_1$, which are not equal in value and angle-dependent. As discussed above, this is due to the loss of e-h symmetry, not described appropriately by the continuum approximation. Therefore, we consider them as parameters given by the TB calculation.  

Finally, the effect of an external applied bias $V$ between the outer layers 1 and 3 is described by adding de corresponding diagonal terms to the effective Hamiltonian, Eq. \ref{hameff}.  
 This does not open a gap in the Dirac cone, but produces a shift in the Dirac point and modifies the gap between the parabolic bands. As an example, adding $-V/2$ to the first diagonal element and $+V/2$ to the diagonal of the $2\times2$ box  corresponding to the twisted layer, induces a shift  $V/2$ in the Dirac cone and modifies the parabolic gap to $\epsilon_1-V/2$. Thus, the application of a bias may close the gap of the parabolic bands and, for negative $V$, it can shift down the Dirac cone.

\section{Summary and Conclusions}
\label{sec:sum}

In summary, we have shown that the low-energy spectrum of the twisted trilayer graphene composed by two Bernal-stacked layers and a third rotated graphene sheet is a combination of the bilayer Bernal-like parabolic bands and linear Dirac cone reminiscent of the twisted bands. The parabolic bands show a gap that gets larger for smaller angles. There is also a shift in the Dirac cone that has a similar angular dependence. The velocity of the twisted-like bands is renormalized for small angles, tending to zero when the rotation angle tends to zero. We find that the gap is correlated with the shift of the Dirac cone in an isolated twisted bilayer.  The shift is due to the loss of the electron-hole symmetry produced by the mixing of the two sublattices. 

One of the parabolic bands has a non-bonding character, pinned at zero energy at the Dirac point, and its carriers are located exclusively in the outer Bernal layer. The other bands have a carrier distribution that depends on the rotation angle. For larger angles, the parabolic band is mostly located on the inner Bernal layer, whereas the carriers of the linear bands are mainly on the twisted layer. However, for angles below 
 $5^{\rm o}$ there is an increasing weight on the twisted layer for the parabolic band, and the linear bands, initially located on the twisted layer, gain weight in the other two AB-stacked layers. 
 
 The change in localization indicates an increase in the effective coupling between layers, and coincides with a change in the relative position of the two parabolic bands, i. e., a zero gap point. This also concurs with the onset of the velocity renormalization, consistent with the aforementioned enhancement of the effective interlayer interaction. 
 
We have compared our results obtained with a tight-binding model with those derived from a continuum approximation. While the velocity renormalization it is correctly given by the continuum model, the shift of the Dirac cone is not well described by this approximation, neither with respect to the sign nor to the angular dependence. We thus derive an effective model for the low energy physics of the system, allowing for a simpler description of its properties.

\begin{acknowledgments}

This work has been partially supported by MEC-Spain under grant
FIS2012-33521. 
E.S.M. acknowledges DGIP/USM for the internal grant 111217. M. P. thanks 
FONDECYT grant 1100672 and DGIP/USM internal grant  11.11.62. The authors thank H. Santos,   J.D. Correa and P. San-Jose for helpful discussions. 

\end{acknowledgments} 

\appendix
\section{Continuum model for the twisted trilayer graphene}
We consider a twisted trilayer graphene formed by three graphene monolayers, with two of them ($1$ and $2$) $AB$-stacked, and the third one rotated a commensurate angle $\theta$ about a perpendicular axis passing through 
a $B2$ atom (Fig.  \ref{fig:geom}). 
Using a superlattice unit cell defined by the basis vectors ${\bf t}_1 = (i+1) {\bf a}_1+i {\bf a} _2 $ and  ${\bf t}_2 = - i {\bf a}_1+ (2i+1) {\bf a} _2 $, where $i$ is a positive integer, we get a commensurate rotation angle defined by 
\begin{equation}
\cos \theta = \frac {3 i^2+3i+1/2}{3i^2+3i+1} \, .
\end{equation}

We are interested in the low energy spectrum near a Dirac point, ${\bf K}$=$\frac{4\pi}{3a_0}(1,0)$, in layer $1$ and $2$. Under the rotation, the equivalent Dirac point in layer $3$ is 
${\bf K} ^{\theta}= \frac{4\pi}{3a_0}(\cos \theta, \sin\theta)$.   Setting $\Delta {\bf K}$=${\bf K} ^{\theta}-{\bf K}$, we can define wave functions with the same wavevector ${\bf k}$  in the three layers that refer to the same plane wave states in the original lattice: the Dirac points occur at $-\Delta {\bf K}/2$ in layers $1$ and $2$ and at $\Delta {\bf K}/2$ in layer $3$.
With this the intralayer part of the Hamiltonian takes the form
\begin{eqnarray}
&&
H_{c\,1}+H_{c\,2}+H_{c\,3} = 
v_F \sum _{{\bm k},\alpha,\beta} c^+ _{1,{\bf k},\alpha} {\bm \sigma} _{\alpha,\beta} \cdot \left (  {\bf k}+ \Delta {\bf K}/2 \right ) c _{1,{\bf k},\beta}  
\nonumber \\
&&+v_F \sum _{{\bf k},\alpha,\beta} c^+ _{2,{\bf k},\alpha} {\bm \sigma} _{\alpha,\beta} \cdot  \left (  {\bf k}+ \Delta {\bf K}/2 \right ) c _{2,{\bf k},\beta} \nonumber \\ &&+
v_F \sum _{{\bf k},\alpha,\beta} c^+ _{3,{\bf k},\alpha} {\bm \sigma} ^{\theta}_{\alpha,\beta} \cdot \left (  {\bf k}-\Delta {\bf K}/2 \right ) c _{3,{\bf k},\beta} \, , 
\label{hintra}
\end{eqnarray}
with ${\bm \sigma} ^{\theta}= e ^{i \theta \sigma _z /2} (\sigma _x,\sigma _y) e ^{-i \theta \sigma _z /2}$.  In this expression the operator $ c^+ _{i,{\bf k},\alpha}$ creates 
and electron in layer $i$ on atom $\alpha$ and momentum ${\bf k}$ and $v_F$=$\sqrt{3}/2 \gamma_0 a_0$.

The tunneling between layers $1$ and $2$, which are 
$AB$-stacked, 
is described by the interlayer 
Hamiltonian
\begin{equation}
H_{c\,12} = \sum _{\alpha, \beta,{\bf k}} \left (t^{AB} _{\alpha,\beta}   c^+ _{1,{\bf k},\alpha}c _{2,{\bf k},\beta}  +H.c. \right ) ,
\label{tun12}
\end{equation}
where the hopping matrix 
for the $AB$ stacking 
is 
\begin{equation}
t^{AB} _{\alpha,\beta} = \gamma_1 \left (   \begin{array}{cc}
    1 & 0 \\ 
    0 & 0\\ 
  \end{array}
\right ) .
\end{equation}
We describe the coupling between the twisted layers 2 and 3 with the approximation 
introduced by Lopes dos Santos.\cite{Lopes_2007} We consider a periodically modulated local tunneling 
\begin{equation}
H_{c\,23} = \sum _{\alpha, \beta,{\bf k},{\bf G}} \left (t^{TW} _{\alpha,\beta} ({\bf G})  c^+ _{2,{\bf k}+{\bf G},\alpha}c _{3,{\bf k},\beta}  +H.c. \right ) ,
\label{tun23}
\end{equation}
where the vector ${\bf G}$ is summed over the reciprocal lattice vectors of the supercell. In the continuum approximation, only the largest
$t^{TW} _{\alpha,\beta} ({\bf G})$ are 
retained in the hopping Hamiltonian. They 
correspond to the reciprocal lattice vectors
${\bf 0}$, ${\bf G}_1$ and ${-\bf G} _{1} -{\bf G} _2$  and have the form 
\begin{eqnarray}
t^{TW} _{\alpha,\beta}(0) & = & \tilde{\gamma} \left (   \begin{array}{cc}
    1 & 1\\ 
    1 & 1\\ 
  \end{array}
\right )  , \,\,
t^{TW} _{\alpha,\beta}({\bf G}_1) = \tilde{\gamma} \left (   \begin{array}{cc}
    f& f^*\\ 
    1 & f\\ 
  \end{array}
\right )  , \nonumber \\
t^{TW} _{\alpha,\beta}(&- & {\bf G}_1  -  {\bf G}_2) = \tilde{\gamma} \left (   \begin{array}{cc}
    f^* & f\\ 
    1 & f ^*\\ 
  \end{array}
\right ),  
\end{eqnarray}
with 
$f=e^{i2\pi/3}$ and $ \tilde{\gamma}$=$\gamma_1/3$. In the previous expressions ${\bf G}_1$ and ${\bf G}_2$ are the primitive vectors of the reciprocal space  defined by the supercell lattice vectors
${\bf t}_1$ and ${\bf t}_2$.

The band structure is obtained by diagonalizing the Hamiltonian
\begin{equation}
H_c = H_{c\,1}+H_{c\,2}+H_{c\,3}+H_{c\,12}+H_{c\,13}
\, \, 
\end{equation}
in a plane wave basis. The number of plane waves needed 
to obtain an accurate 
spectrum increases with the size of the supercell, and therefore the 
calculation is more 
demanding numerically 
for smaller
twisted angles.  We have checked the convergence of our results with respect to the number of plane waves used.
\begin{figure}
\includegraphics[width=8cm]{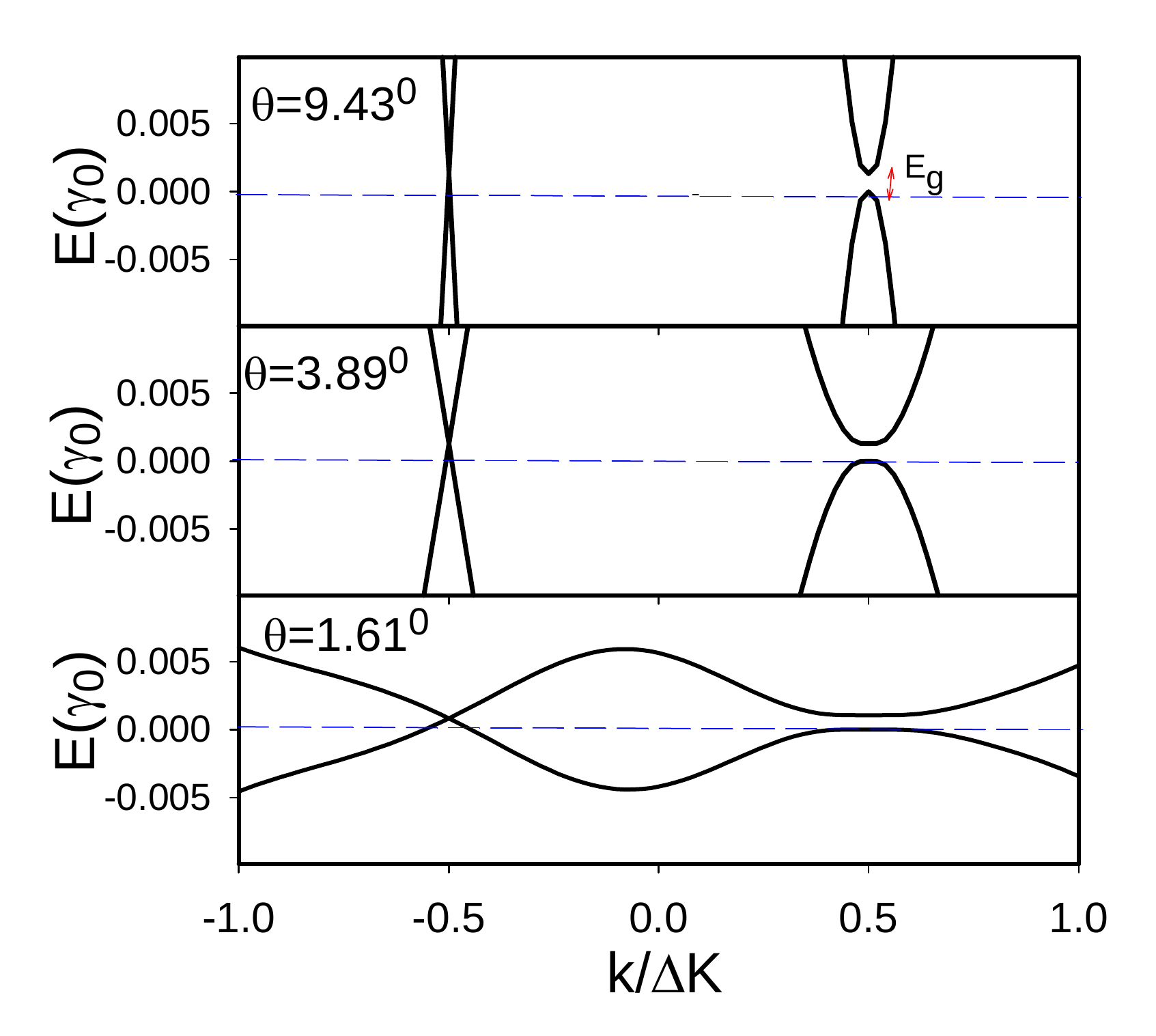}
\caption{(Color online) Band structure of twisted trilayer graphene as obtained in the continuum approximation. In the calculation $\gamma_1$=0.12$\gamma_0$. } 
\label{bandas_cont}
\end{figure}
In Fig. \ref{bandas_cont} we plot the low energy bands, as obtained by diagonalizing the continuum Hamiltonian for three values of the twisted angle.  The spectrum consists of
a
Dirac cone located mainly on the twisted layer and two gapped parabolic bands  with states 
 in the $AB$-stacked bilayer.
 The continuum results are the following: 
(i) the velocity of the Dirac cone $\tilde v$ and the curvature of the parabolic bands decrease 
with diminishing rotation angle; 
(ii) there 
is 
an energy shift of the Dirac cone, $\bar {\epsilon} _0$, equal to the gap between the parabolic bands;  
(iii) the top of the parabolic valence band is pinned at zero energy;  
(iv) the shift and gap decrease when the twisted angle decreases. 
We have checked that $\bar {\epsilon} _0$ and $\tilde v$ coincide with the energy shift and the velocity renormalization occurring in twisted bilayer graphene.\cite{Lopes_2007}
From the previous observations we can define a Hamiltonian 
which describes 
the low energy physics of twisted trilayer graphene for angles between 1$^{\rm o}$ and 20$^{\rm o}$,
\begin{equation}
H^{6\times 6}=  \left( \begin{array}{cccccc}
0 & v_F g_+  &\gamma_1 &0 &0 &0 \\
v_F g^* _+ &0 &0 &0 &0 &0 \\
\gamma_1 & 0 & \bar {\epsilon} _0& \tilde v g _+  &0 &0 \\
0& 0 & \tilde v  g^*  _+  &\bar {\epsilon} _0& 0&0\\
0& 0 &0& 0& \bar {\epsilon} _0&  \tilde v  g _-  \\
0& 0 &0& 0& \tilde v    g^* _- & \bar {\epsilon} _0\\
  \end{array} \right),
\end{equation} 

\noindent
where  $g _{\pm}$=$g({\bf k}\pm  \frac {\Delta {\bf K}} 2) $ and  $g(\bf k)$=$k_x -i k_y $. This Hamiltonian can be further simplified by eliminating the dimer states in the $AB$ stacked layers,\cite{McCann_2006} 
\begin{equation}
H^{4  \times 4}=  \left( \begin{array}{cccc}
0 &  \frac {v_F \tilde v }{\gamma _1}  (g_+)^ 2 &0 &0 \\
\frac {v_F \tilde v }{\gamma _1}     (g^* _+)^2  &\bar {\epsilon} _0&0 &0  \\
0& 0& \bar {\epsilon} _0&  \tilde v   g_-   \\
0& 0& \tilde v    g_-^*   & \bar {\epsilon} _0\\
 \end{array} \right) .
\end{equation}


\begin{thebibliography}{28}
\bibitem [{not()}]{notahop13}%
  \BibitemOpen
  \href@noop {} {}\bibinfo {note} {We have checked that the inclusion of
  interlayer couplings between the two outer layers only modifies the numerical
  values of the band energies, mainly the shift of the Dirac cone.}\BibitemShut
  {Stop}%
\end{thebibliography}
\end{document}